\documentclass[reprint,amssymb,amsmath,aip,cha]{revtex4-1}
\usepackage{epsfig}
\usepackage{dcolumn}
\usepackage{bm}

\begin{document}

\title{Phase response curves for models of earthquake fault dynamics}

\author{Igor Franovi\'c}%
\email{franovic@ipb.ac.rs}%
\affiliation{Scientific Computing Laboratory, Institute of Physics Belgrade, University of Belgrade,
Pregrevica 118, 11080 Belgrade, Serbia}%

\author{Srdjan Kosti\'c}%
\affiliation{Institute for the Development of Water Resources "Jaroslav \v{C}erni",
Jaroslava \v{C}ernog 80, 11226 Belgrade, Serbia}%

\author{Matja\v{z} Perc}%
\affiliation{Faculty of Natural Sciences and Mathematics, University of Maribor, Koro{\v s}ka cesta 160, SI-2000 Maribor, Slovenia}%
\affiliation{CAMTP -- Center for Applied Mathematics and Theoretical Physics, University of Maribor, Krekova 2, SI-2000 Maribor, Slovenia}%

\author{Vladimir Klinshov}%
\affiliation{Institute of Applied Physics of the Russian Academy of Sciences, 46 Ulyanov Street, 603950 Nizhny Novgorod, Russia}%

\author{Vladimir Nekorkin}%
\affiliation{Institute of Applied Physics of the Russian Academy of Sciences,
46 Ulyanov Street, 603950 Nizhny Novgorod, Russia}%
\affiliation{University of Nizhny Novgorod, 23 Prospekt Gagarina, 603950 Nizhny Novgorod, Russia}%

\author{J\"urgen Kurths}%
\affiliation{Institute of Applied Physics of the Russian Academy of Sciences,
46 Ulyanov Street, 603950 Nizhny Novgorod, Russia}%
\affiliation{Potsdam Institute for Climate Impact Research, 14412 Potsdam, Germany}%
\affiliation{Institute of Physics, Humboldt University Berlin, 12489 Berlin, Germany}%

\date{\today}

\begin{abstract}
We systematically study effects of external perturbations on models describing earthquake fault dynamics. The latter are based on the framework of the Burridge-Knopoff spring-block system, including the cases of a simple mono-block fault, as well as the paradigmatic complex faults made up of two identical or distinct blocks. The blocks exhibit relaxation oscillations, which are representative for the stick-slip behavior typical for earthquake dynamics. Our analysis is carried out by determining the phase response curves of first and second order. For a mono-block fault, we consider the impact of a single and two successive pulse perturbations, further demonstrating how the profile of phase response curves depends on the fault parameters. For a homogeneous two-block fault, our focus is on the scenario where each of the blocks is influenced by a single pulse, whereas for heterogeneous faults, we analyze how the response of the system depends on whether the stimulus is applied to the block having a shorter or a longer oscillation period.
\end{abstract}

\pacs{05.45.Xt, 91.30.Ab, 02.30Ks}

\maketitle

\begin{quotation}
Earthquakes are conceptually considered as frictional instabilities occurring on preexisting tectonic faults. The fault dynamics is often represented by the class of spring-slider block models incorporating different forms of constitutive friction laws. Such models can qualitatively account for the relevant regimes of fault dynamics, including the aseismic creep motion or the stick-slip motion, which is a signature for earthquakes. The research involving these models has so far mainly aspired to gain insight into the scaling laws and the occurrence of characteristic events, as well as to elucidate the relation between small and large earthquakes. Here we adopt a considerably different approach. Our intention is not to characterize the statistical properties of the underlying time series or to assess the earthquake hazard, but rather to analyze a representative class of fault models from the perspective of nonlinear dynamics theory. Being strongly nonlinear systems, the considered models of earthquake faults are expected to display a number of intricate features, including high sensitivity to external perturbation, whereby the response may qualitatively depend on different system parameters, as well as the fault complexity. In the present paper, the theory of phase response curves is applied for the first time to systematically examine the sensitivity of fault dynamics in the stick-slip regime to external perturbation. We consider the cases of a simple monoblock and paradigmatic two-block complex faults. Perturbations made up of single or two consecutive pulses are found to affect the fault dynamics in a nontrivial fashion, being able to advance or delay the earthquake cycle or even give rise to long-term effects.
\end{quotation}

\section{Introduction}{\label{sec:intro}

By a basic phenomenological description, earthquakes are regarded as large-scale recurring mechanical failure events \cite{KHKBC12}, characterized by seismic cycles comprised of a comparably long quasi-static stage of tectonic stress build-up and an abrupt dynamical rupture stage, associated with a rapid release of the accumulated strain. Earthquakes occur as dynamical instabilities on preexisting crustal faults and are caused by the motion of tectonic plates, which is fundamentally influenced by the elastic properties of the crust and the frictional features of the fault \cite{S02,K09}. In dynamical terms, the complexity in earthquake-related behavior derives from the coaction of intrinsic nonlinearity, dominated by friction, and the external driving. Typically, the physical background behind inter-plate earthquakes involves a fault segment, represented by a mass block or an assembly of blocks, which loaded by one tectonic plate and under the frictional resistance of the other plate exhibits a stick-slip behavior \cite{KHKBC12,S02}, the type of motion paradigmatic for earthquake dynamics \cite{S02,BB66,BLH01}. In different physical models, the friction terms attain quite a complex form and are expressed by appropriate "constitutive laws" \cite{S02,K09,M98}.

Apart from scale-invariant statistical features, reflected in several well-known empirical scaling laws, some earthquakes exhibit characteristic features, manifested as "characteristic earthquakes" with well-defined time or energy scales \cite{S02,K09,BC06}. In the former case, the fault fails in a pseudo-periodic time series, such that its dynamics is reminiscent of a relaxation oscillator. The pertaining oscillations may naturally be sensitive to external perturbations, which can be derived from different kinds of additional forcing whose duration and magnitude are small compared to the tectonic load. In general, if the perturbation is sufficiently small, it does not influence the amplitude of oscillations, but it may considerably affect the phase. In conceptual terms, and especially from the seismological point of view, it becomes relevant to determine whether and how the phase of stick-slip oscillations, and thereby the characteristic event itself, is retarded or advanced by such perturbations.

In the present paper, we consider the sensitivity to external perturbation of the models of a simple mono-block fault and a paradigmatic two-block complex fault, which display relaxation oscillations with the signature stick-slip property. The models are formulated within the Burridge-Knopoff framework of coupled spring-block systems \cite{BK67,CL89,CL91,C91}, and incorporate the Dieterich-Ruina rate- and state-dependent friction law \cite{R83,RT86,RLR01}. The qualitative analysis will be focused on determining the first- and second-order phase response curves (phase resetting curves, $PRC$s) for these models \cite{K03,W80,T07,SPB12,C06}, which to our knowledge is the first time that such an approach is applied in the context of earthquake fault dynamics, despite
the fact that the formalism related to phase description of nonlinear oscillators has already been invoked \cite{S02,S10,SHK14,LPG10,VCW15}. So far, the $PRC$s have often been used as a tool to study the system's response to stimuli, as well as the units' ability to synchronize in the fields of neuroscience \cite{PV07,TR07,KBP13,AC09} and the general theory of coupled phase oscillators \cite{KNAKK08,KKNAK09,KE09,LP10}.

The main corpus of issues we address here includes $(i)$ the sensitivity of a simple monoblock fault to external perturbation, $(ii)$ the influence of system parameters on the profile of $PRC$s, $(iii)$ the effect of
two-pulse perturbations and the deviation from the superposition principle due to multidimensionality of
the model, as well as $(iv)$ the responses of compound faults, either homogeneous or heterogeneous, to
external perturbation. Apart from considering the first-order $PRC$s, our interest will also lie with the second-order $PRC$s because their nontrivial behavior may indicate a potential long-term effect of external perturbation on the duration of an earthquake cycle. The research agenda has a systematic character precisely given the fact that this type of analysis has not been carried out before for models of fault dynamics.

As already mentioned, the pseudo-periodic recurrence times have primarily been associated with large characteristic earthquakes \cite{W94,S96,TRWML01,NB87,P08,VCW15}. By one scenario, the latter involve breaking of the most part of or the entire seismogenic zone \cite{S02,K09,SM06,I04}, whereas by the other scenario they emerge due to breaking of similar sections of complex faults \cite{S02,B13}. Well-known examples are earthquakes in the Nankaido region (Japan), the northern, southern and Parkfield sections of the San Andreas Fault \cite{SBWF10}, and several regions in China \cite{KETN11,HWTK05}. Apart from these large characteristic earthquakes, the description of fault dynamics in terms of relaxation oscillator models may further be justified for certain small repeating earthquakes \cite{MIH02}, as corroborated by the recent proxy data \cite{NM97}. One should note that many of the relevant models of fault dynamics may yield periodic sequences of events or series with a strong periodic component. For instance, such behavior has been found for the one- and two-dimensional versions of the Burridge-Knopoff model \cite{CL91,C91,MK05,MK06}, as well as in case of the Olami-Feder-Christensen model \cite{KYK08}. It has also been indicated that models of coupled relaxation oscillators displaying the stick-slip dynamics may account for a phase-locking mechanism behind earthquake clusters. The latter conform to rupture patterns where the main events occur in groups comprising nearby or distributed faults with similar characteristic periods \cite{SS13,S10,SHK14}.

The paper is organized as follows. In Sec. \ref{sec:mod}, we introduce the models of a simple fault and a two-block  complex fault, summarizing the results of bifurcation analysis and explaining the physical background and possible regimes of system behavior. Section \ref{sec:single} concerns the monoblock fault, considering the scenarios where the fault is subjected to a single or two successive excitations. In the latter case, we demonstrate a nonlinear effect which occurs for systems whose dimension is larger than $1$ and consists in a deviation from the superposition principle for two subsequent perturbations. It is also discussed how sensitivity to perturbation depends on the system parameters. Section \ref{sec:compfaults} provides our results for the first- and second-order phase response curves in cases of the homogeneous and the heterogeneous two-block complex fault. For homogeneous fault, we analyze how the system responds in case where each of the blocks is perturbed, but the perturbations arrive with a certain phase lag. For the heterogeneous fault, it is examined how the system response changes depending on whether the block with a shorter or longer oscillation period is perturbed. Section \ref{sec:summ} contains a brief summary of our results.

\begin{figure}
\centering
\includegraphics[scale=0.33]{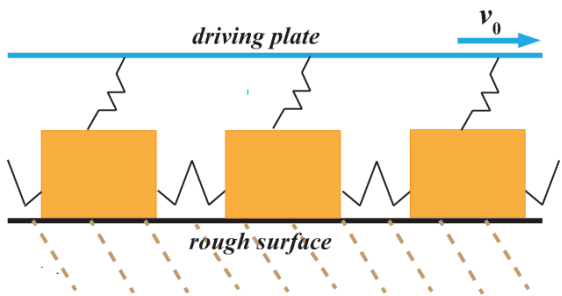}
\caption{(Color online) Schematic representation of the spring-block model of earthquake fault dynamics. The blocks interact via elastic springs, and each block is further elastically coupled to the loader plate which moves at a constant velocity $v_0$. The blocks slide over a rough surface, whereby the friction at their interface is typically described by complex constitutive laws. \label{Fig1}}
\end{figure}

\section{Model of fault dynamics} \label{sec:mod}

Within the family of spring-block models, the fault dynamics is represented by elastically interacting mass blocks sliding over a rough surface, whereby each block is elastically coupled to a rigid loader plate that moves at a constant velocity, see Fig. \ref{Fig1}. In terms of seismological interpretation, it is the interface between the slider blocks and the rough surface that can be considered as an analogue for a one-dimensional earthquake fault \cite{CLST91}, and one is interested in describing the sliders' slip and the associated slip velocity relative to the loader plate. The model comprising a single block accounts for a simple fault, whereas models containing multiple blocks refer to multi-segment (complex) faults. In the present paper, we study the cases of a simple fault and the paradigmatic two-block complex fault, made up of homogeneous or heterogeneous blocks. The block dynamics is provided by a version of the Burridge- Knopoff model supplied by the Dieterich-Ruina rate- and state-dependent friction law \cite{ EBL08,EBL11,KFPVT14,KVFT13,KFTV13}. Note that the selection of friction law is an important point for the models where friction enters as a force term. The early friction laws included effects of slip-weakening (reduction of friction strength during sliding) and rate-weakening (reduction in frictional force which accompanies the increase in slip velocity) \cite{R83}, but the former could not fully explain for the relationship between the static and dynamic friction, while experimental data have further shown that friction could not be a function dependent only on velocity \cite{M98}. The Dieterich-Ruina law \cite{R83,RT86,RLR01} resolves these issues by introducing an additional state variable, which may be attributed a microscopic interpretation, associating it to the average life time of asperity contacts at the interface between the blocks and the rough surface \cite{S02}.

Without specifying the details of the derivation, which can be found in \cite{EBL08,EBL11}, here we provide the final non-dimensional form of equations for the dynamics of a single block:
\begin{align}
\frac{d\theta}{dt}&=-v(\theta+(1+\epsilon)\ln v) \nonumber\\
\frac{du}{dt}&=v-1 \nonumber\\
\frac{dv}{dt}&=-\gamma^2(u+(1/\xi)(\theta+\ln v)).\label{eq1}
\end{align}
In eq. \eqref{eq1}, $\theta$ represents the state variable, whereas $u$ denotes the slip (relative to the driver plate) and $v$ is the associated slip velocity. The strong nonlinearity of \eqref{eq1} is due to the friction term, which involves a logarithm dependence on the velocity. The parameters $\xi$ and $\gamma$ are the non-dimensional spring constant and the non-dimensional frequency, respectively. The spring stiffness qualitatively accounts for the elastic properties of the medium where the fault is embedded \cite{S02}. The parameter $\epsilon$ essentially measures the sensitivity of the block's velocity relaxation. This interpretation derives from the point that $\epsilon$ may be expressed via two additional stress parameters related to the velocity dependence on the friction stress $\tau$. In particular, $\epsilon$ is given by the ratio $\epsilon=(B-A)/A$ \cite{EBL08,EBL11}, where $A$ presents the direct velocity dependence $A=\frac{\partial \tau}{\partial ln(v)}$, while $A-B=\frac{\partial \tau_{ss}}{\partial \ln(v_{ss})}$ is the velocity dependence for the steady state \cite{R83,RLR01}, when the slider moves at a constant velocity $v_{ss}$. In other words, $\epsilon$ is determined by the ratio of stress dropped during the earthquake to the stress increase that accompanies a sudden change in the block velocity. Note that we consider only positive values of $B-A$ ($\epsilon>0$), which from a micro-mechanical point of view corresponds to the velocity-weakening effect \cite{M98}. Compared to real fault conditions, $A$ and $B$ describe material properties that depend on pressure, temperature and sliding velocity \cite{S02}. These arguments suggest that $\epsilon$ is the parameter most specific to detailed dynamics of particular faults. In terms of a qualitative comparison to real earthquake faults, it has been established that the relevant range of values for the parameters $\epsilon,\xi$ and $\gamma$ is $\epsilon\in(1,3.5),\xi\approx0.5,\gamma\in(10^3-10^{12})$ \cite{RT86,EBL08,BMLBK98}.

\begin{figure*}
\centerline{\epsfig{file=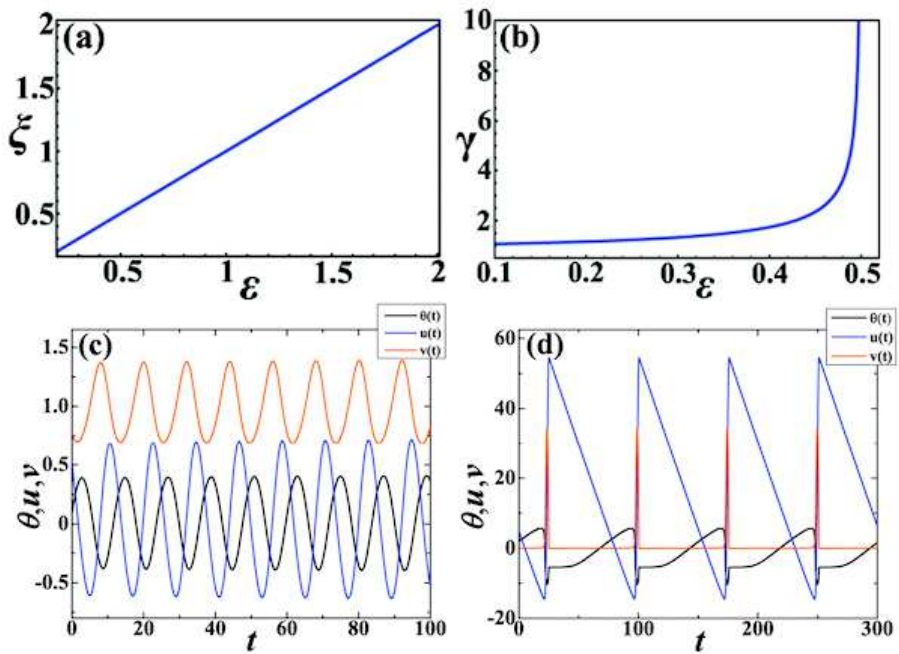,width=12cm}}
\caption{(Color online) Bifurcation diagrams and characteristic regimes of motion for the system \eqref{eq1}. In (a) is shown the Hopf bifurcation curve $\xi(\epsilon)$ obtained for fixed $\gamma=1000$. In (b) is presented the Hopf bifurcation curve $\gamma(\epsilon)$ determined for fixed $\xi=0.5$. (c) and (d) illustrate the dynamics associated with the creep regime (harmonic oscillations) and the stick-slip regime (relaxation oscillations), respectively. (c) is obtained for the parameter set $(\epsilon,\xi,\gamma)=(0.3,0.5,0.8)$, whereas the parameters in (d) are $(\epsilon,\xi,\gamma)=(1.45,0.5,1000)$. \label{Fig2}}
\end{figure*}

System \eqref{eq1} has a stationary state $(\theta,u,v)=(0,0,1)$ which corresponds to sliding at a uniform velocity equal to that of the loader plate, such that the block exhibits no slip relative to the plate. For certain parameter values, the system undergoes a direct supercritical Hopf bifurcation which gives rise to an oscillatory solution. The corresponding bifurcation curves $\xi(\epsilon)$ obtained for fixed $\gamma$ and $\gamma(\epsilon)$ under fixed $\xi$ are shown in Fig. \ref{Fig2}(a) and \ref{Fig2}(b). Note that these curves are determined analytically by considering the pure imaginary roots of the characteristic equation $\lambda^3-\lambda^2(\frac{\gamma^2}{\xi}-1)-\lambda\gamma^2(1-\frac{\epsilon}{\xi})-\gamma^2=0$
for system \eqref{eq1}. Immediately above the bifurcation curves, system \eqref{eq1} displays harmonic oscillations, cf. Fig. \ref{Fig2}(c), which may be appropriate to describe pre-seismic and post-seismic creep regimes \cite{KFPVT14}. Nevertheless, the regime of relaxation oscillations which we are interested in, can be found sufficiently away from criticality, see Fig. \ref{Fig2}(d). Such relaxation oscillations can be considered as dynamical counterpart of the stick-slip behavior paradigmatic for earthquake motion. In the quasi-static stage of stress accumulation (the "stick" stage), the block is effectively stuck on the rough surface, so that the relative slip to the driver plate decreases at a constant rate as the driver plate first catches up and then even surpasses the block. Once the pulling force overcomes the static friction withholding the block, one arrives at the onset of the slip stage. At this point, the block's velocity increases sharply, such that the slider shoots forward again, which gives rise to a new seismic cycle.

Apart from the monoblock fault, we also consider the case of a two-component fault, where the dynamics of blocks is given by:
\begin{align}
\frac{d\theta_i}{dt}&=-v_i(\theta_i+(1+\epsilon_i)\ln v_i) \nonumber\\
\frac{du_i}{dt}&=v_i-1 \nonumber\\
\frac{dv_i}{dt}&=\gamma_i^2(c(u_i-u_j)+u_i+(1/\xi_i)(\theta_i+\ln v_i)),\label{eq2}
\end{align}
with $i,j\in\{1,2\},i\neq j$. The interactions are characterized by the coupling strength $c$. We intend to analyze the sensitivity to perturbation of the homogeneous two block fault ($\epsilon_1=\epsilon_2, \xi_1=\xi_2,\gamma_1=\gamma_2$), as well as the heterogeneous complex fault. Consistent with the arguments regarding the system parameters, heterogeneity will be confined to the case of two blocks with disparate $\epsilon$, $\epsilon_1\neq \epsilon_2$, which results in distinct periods of the respective stick-slip oscillations. Both for the homogeneous and the heterogeneous two-segment faults, we take the coupling strength $c$ sufficiently weak so that the interaction does not perturb the respective oscillation cycles of the blocks.

A remark is required regarding the numerical treatment of models \eqref{eq1} and \eqref{eq2}. In particular, the underlying systems of $ODE$s are stiff, in a sense that within the relevant parameter domain, an exceedingly small iteration step is required to maintain the numerical stability of the typical explicit integration schemes, such as the Runge-Kutta method. The stiffness feature derives from the fact that the system involves characteristic time scales of substantially different order, and becomes stronger as $\gamma$ is increased. Note that the step size is limited more severely by the stability rather than the accuracy requirement of the integration methods. In order to resolve these issues, we have implemented the solver based on the Rosenbrock method, which is specifically adapted to stiff systems. Unless stated otherwise, the parameter set used for the block dynamics throughout the paper is $(\epsilon,\xi,\gamma)=(1.45,0.5,1000).$

\section{$PRC$s for the monoblock fault}\label{sec:single}

\subsection{Theoretical background} \label{subsec:theory}

Phase response curve is an inherent feature of an arbitrary oscillator, which reflects its sensitivity to a brief (pulse-like) stimulus. $PRC$ is given by the dependence of the phase shift, induced by a perturbation, as a function of the oscillation phase at which the perturbation has occurred. The effect of phase resetting due to pulse perturbation may formally be treated as follows. We first consider a one-dimensional oscillator, described only by a continuously increasing phase variable $\phi$ that evolves as $\dot{\phi}=\omega$. Then the system's phase just after a pulse stimulus of strength $\kappa$ arrived at the moment $t_p$ can be written as \cite{SPB12,KBP13}
\begin{equation}
\phi_{+}(t_p)=\phi(t_p)+\kappa Z(\phi,\kappa), \label{eq3}
\end{equation}
where $Z(\phi,\kappa)$ stands for the $PRC$. In a more general case, periodic oscillations are characterized by a limit-cycle attractor $X_0(t)$ in an $N$-dimensional phase space. Nevertheless, the notion of isochrones \cite{G75,SP13} still allows one to consider a phase space representation of the form ($\textbf{a},\phi$), where $\textbf{a}$ is an $(N-1)$ dimensional "amplitude", and $\phi$ is the regular phase variable obeying $\dot{\phi}=\omega$ \cite{KBP13}. Without loss of generality, one may assume that the "amplitude" vanishes on the limit cycle ($\textbf{a}=0$). In this setup, if a kick is introduced at the moment $t_p$, the reset of the state ($\textbf{a},\phi$) just after $t_p$ may be expressed as \cite{KBP13}
\begin{align}
\textbf{a}_{+}(t_p)&=\textbf{a}(t_p)+
\kappa A(\textbf{a}(t_p),\phi(t_p),\kappa)=\kappa A(0,\phi(t_p),\kappa) \nonumber \\
\phi_{+}(t_p)&=\phi(t_p)+\kappa \Phi(\textbf{a}(t_p),\phi(t_p),\kappa)=\phi(t_p)+ \kappa Z(\phi(t_p),\kappa). \label{eq4}
\end{align}
The above equations take into account that the initial state lies on the limit cycle ($\textbf{a}=0$), such that $\Phi(0,\phi,\kappa)=Z(\phi,\kappa)$ holds. In terms of application, $PRC$s were first introduced in the study
of oscillations in biological systems, including cardiac cells, fireflies populations and especially neural networks \cite{MS90,EK91,K91,B95,G01,GNHTN02,CA10,KN13}. Within these fields, as well the general theory of coupled phase oscillators, the method has facilitated an analysis of the units' interaction properties, including stability, synchronization or clustering. The concept of $PRC$s allows one to reduce complex models of oscillators to simple phase models which still reflect important features of the original oscillators, viz. the point that the effect of perturbation depends on the dynamical state of the oscillator. In its representation as a phase oscillator, each oscillator possesses a characteristic $PRC$ that can be computed numerically or measured experimentally
\cite{E96,EBTN11,GEU05,KSN14,PLV15}.

Let us now address the details relevant for obtaining the $PRC$s in case of our models of earthquake fault dynamics. In a general multidimensional system, the kick may be applied to any of the system variables. Here, a perturbation is added to the second equation of the system \eqref{eq1} or \eqref{eq2}, which is the most plausible choice, because it may be interpreted as a small variation at the loading point. The corresponding equation then takes the form $\frac{du}{dt}=v-1+f(t)$, where $f(t)$ is the perturbation term. In real faults, such perturbations may derive from various natural and artificial sources, including rock break, pressure fluctuations or crack vibration due to movement of magma and volcanic gases \cite{C96}, sudden stress drops \cite{B03,P03}, drilling and blasting in underground mining activities \cite{L12,MB14}, as well as microearthquakes due to hydraulic fracturing or deep injection of waste fluids \cite{E13}.

\begin{figure}
\centering
\includegraphics[scale=0.345]{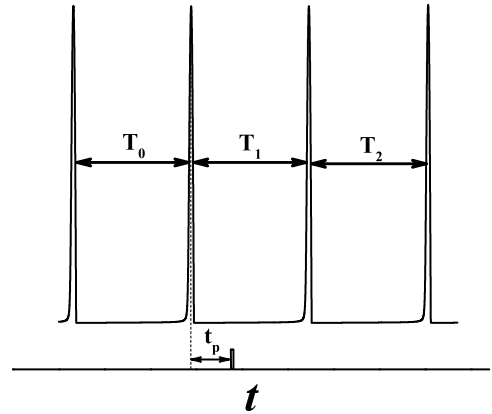}
\caption{Illustration of the method used to determine the $PRC$s. The method is based on measuring the perturbation induced changes in oscillation periods of cycles where the pulse perturbation arrived (first-order $PRC$) and the next oscillation cycle (second-order $PRC$). $T_0$ denotes the default oscillation period, $T_1$ is the duration of the oscillation cycle influenced by the pulse at phase $\phi=t_p/T_0$, whereas $T_2$ is the duration of the subsequent cycle. \label{Fig3}}
\end{figure}

In order to determine the $PRC$s, one does not have to carry out an explicit phase reduction of the underlying systems, but may rather focus on the occurrence of characteristic events. The latter are associated to large spikes of block's velocity and are representative of earthquakes within the given models. Then, the $PRC$s may effectively be determined in complete analogy to the method typically used for systems of spiking neurons. In particular, the impact of a perturbation is such that it locally changes the oscillation period of a system from the default value $T_0$ (period in the absence of perturbation) to a different value $T_1$, see Fig. \ref{Fig3}. One may use this to numerically determine the phase shift $\Delta\phi$ by measuring the relative change of the period
\cite{T07,SPB12,KBP13,ET10,KSN14}
\begin{equation}
\Delta\phi(\phi)=\frac{T_0-T_1}{T_0}. \label{eq5}
\end{equation}
The phase shift $\Delta\phi$ plotted as a function of the phase $\phi$ when a perturbation has kicked in is precisely the $PRC$. If $T_1<T_0$, the stimulus advances the cycle and \emph{vice versa}. The change of period of the oscillation cycle where a perturbation has occurred defines the first-order $PRC$. Perturbations may also affect the duration of the next oscillation cycle $T_2$, which corresponds to the second-order $PRC$, where the phase shift is given by
\begin{equation}
\Delta\phi^{(2)}(\phi)=\frac{T_0-T_2}{T_0}. \label{eq51}
\end{equation}
In the seismological context, the second-order $PRC$ may be interpreted as qualitatively accounting for a long-term effect of an external perturbation to the pertaining fault dynamics, bearing in mind that the interseismic periods typically comprise very long time scales.

\begin{figure}
\centering
\includegraphics[scale=0.33]{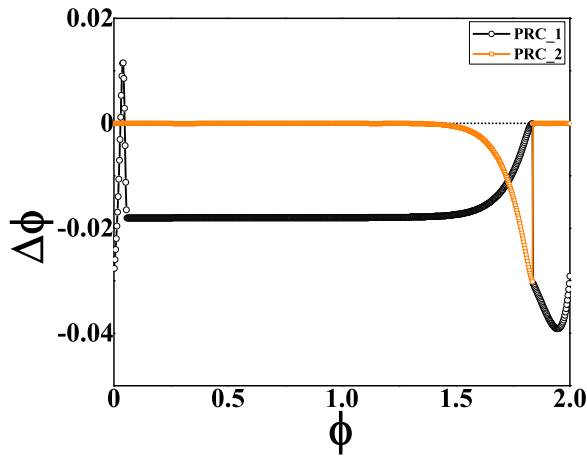}
\caption{ (Color online) $PRC$s of the first (black circles) and second order (orange squares) for a monoblock fault in the stick-slip regime. The system parameters are $(\epsilon,\xi,\gamma)=(1.45,0.5,1000)$. \label{Fig4}}
\end{figure}

By implementing the described method, we determine the first- and second-order $PRC$s for different models of fault dynamics. Apart from a single pulse perturbation, we also consider scenarios where two subsequent pulses are introduced within a given oscillation cycle. The details regarding the validity of the superposition principle in this case will be discussed in Sec. \ref{sec:two_pulses}.

The form of the perturbation involves the standard $\alpha$ function $f(t)=C*[(-1/t_f)*\exp(-(t-t_p)/t_f)+(1/t_r)*\exp(-(t-t_p)/t_r)]\Theta(t-t_p)$, whereby the Heaviside $\Theta$ function is used for shifting along the time-axis. Naturally, the rise and decline characteristic times $t_r$ and $t_f$ are selected so that the perturbation maintains a narrow profile compared to the oscillation period $(t_r=0.15,t_f=0.4)$, whereas $C$ should be kept sufficiently small so that the perturbation does not affect the amplitude of the underlying oscillations ($C=5$). A brief remark regarding the explicit form of perturbation is in order. In view of actual fault dynamics, the perturbation form involving the step-like time dependence may be more realistic \cite{S99,DP11,DP15}. Nevertheless, within the $PRC$ framework, it is well established that the profile of $PRC$s is not qualitatively affected by the particular form of perturbation. In terms of application of $PRC$ theory, the only relevant aspects concern the above conditions on the magnitude and duration of perturbation.

Note that in all the considered instances, zero phase is assigned to the maximum amplitude of the $u$ variable, which is in the seismological interpretation a natural choice, because it corresponds to the occurrence of the characteristic event (earthquake).

\subsection{$PRC$s for a single pulse perturbation}\label{sec:one_pulse}

In this subsection, we numerically determine the single-pulse $PRC$s for a simple mono-block fault in the stick-slip regime, and then consider how the $PRC$ profiles are affected by variation of the fault parameters.

The profiles of the first- and second-order $PRC$s are provided in Fig. \ref{Fig4}. Note that in all the figures throughout the paper, the phase values are expressed in units of $\pi$. An important point regarding Fig. \ref{Fig4} is that the first-order $PRC$ shows a phase advancement only in a narrow phase interval, centered at some small time distance after the earthquake. (Recall that the earthquake event is assigned with $\phi=0$). Nevertheless, the external stimulus introduced at all the other points of the oscillation cycle has a retardation effect, viz. it delays the next characteristic event. The change of sensitivity to a perturbation is expectedly found close to the end of the seismic cycle. In that phase domain, the delay effect is weaker, but the perturbation still cannot advance the cycle. We have verified that the characteristic profile of the $PRC$ does not change under variation of the perturbation amplitude within the relevant range of values.

As one may have expected, the second order $PRC$ corroborates that the perturbation typically has a negligible impact on the duration of the next seismic cycle. Nevertheless, an interesting point concerns the existence of a long-term retardation effect for $\phi\approx 1.8\pi$. Note that this pronounced delay effect occurs precisely in the phase domain where the first-order $PRC$s show a reduced retardation.

Let us now examine how robust are the obtained $PRC$ profiles against variation of the fault parameters. It has already been explained that the parameters $\xi$ and $\gamma$ are less specific to particular faults, so that the impact of their variation may be of less significance compared to the effect of changing $\epsilon$, which is highly specific to particular faults. Still, we note that the $PRC$ profiles from Fig. \ref{Fig4} turn out to be generic, i. e. they remain qualitatively unaffected by changing $\xi$ or $\gamma$ for fixed $\epsilon$. The effects of varying $\epsilon$ under fixed $\xi$ and $\gamma$ are demonstrated in Fig. \ref{Fig5}(a) and \ref{Fig5}(b). These figures refer to phase shifts corresponding to first- and second-order $PRC$s respectively, whereby $\epsilon$ attains values from the seismologically relevant range $\epsilon\in(1,3)$, while $\xi$  and $\gamma$ are fixed at values from Fig. \ref{Fig4}. Naturally, the parameters of pulse perturbation are the same as in Fig. \ref{Fig4}.

\begin{figure}
\centering
\includegraphics[scale=0.42]{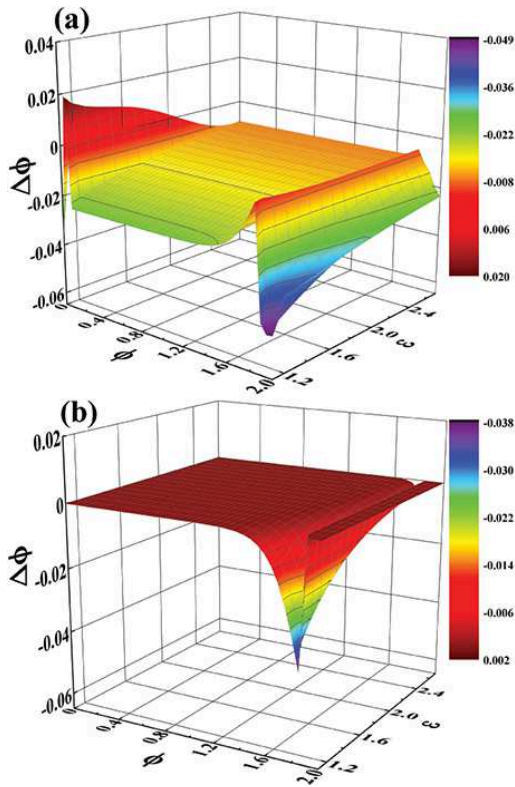}
\caption{(Color online) (a) and (b) respectively show the families of first- and second-order $PRC$s  $\Delta\phi(\phi,\epsilon)$ for a monoblock fault under variation of $\epsilon$. The remaining fault parameters are fixed at $\xi=0.5,\gamma=1000$.
\label{Fig5}}
\end{figure}

Concerning the first-order $PRC$, the effect of advancing the seismic cycle by a perturbation introduced within a preferred time interval just after the earthquake is maintained for most of the considered $\epsilon$ values, but is downgraded with increasing $\epsilon$. In fact, one also finds a critical $\epsilon$ value above which there is no phase advance, cf. Fig. \ref{Fig5}(a). The other interesting effect, which consists in a reduced phase delay if the perturbation occurs close to the end of the seismic cycle, appears unaffected by variation of $\epsilon$. Also, the delay effect characteristic for the most of phase domain is less pronounced with increasing $\epsilon$. Therefore, the profile of the first-order $PRC$ in general becomes more flat as $\epsilon$ is enhanced. A similar statement holds in case of the second-order $PRC$. In fact, Fig.\ref{Fig5}(b) clearly shows that the pronounced delay effect for the perturbation occurring by the end of the oscillation cycle is gradually lost with $\epsilon$.

\subsection{Two-pulse $PRC$s for a monoblock fault}\label{sec:two_pulses}

In this subsection, we consider the response of a monoblock fault in the regime of stick-slip oscillations to two successive pulse perturbations. The first pulse acts at the phase $\phi_1$ of the oscillation cycle, whereas the other pulse is applied at a phase $\phi_2$. Note that the perturbation parameters in both instances are taken to be the same. The occurrence of multiple perturbations during a single oscillation cycle may be attributed to a number of different phenomena, both natural and artificial. Apart from analyzing the pertaining first- and second-order $PRC$s, we also make a remark on the validity of the superposition principle, which assumes the linear summation of phase shifts that result from two successive small perturbations.

\begin{figure}
\centering
\includegraphics[scale=0.56]{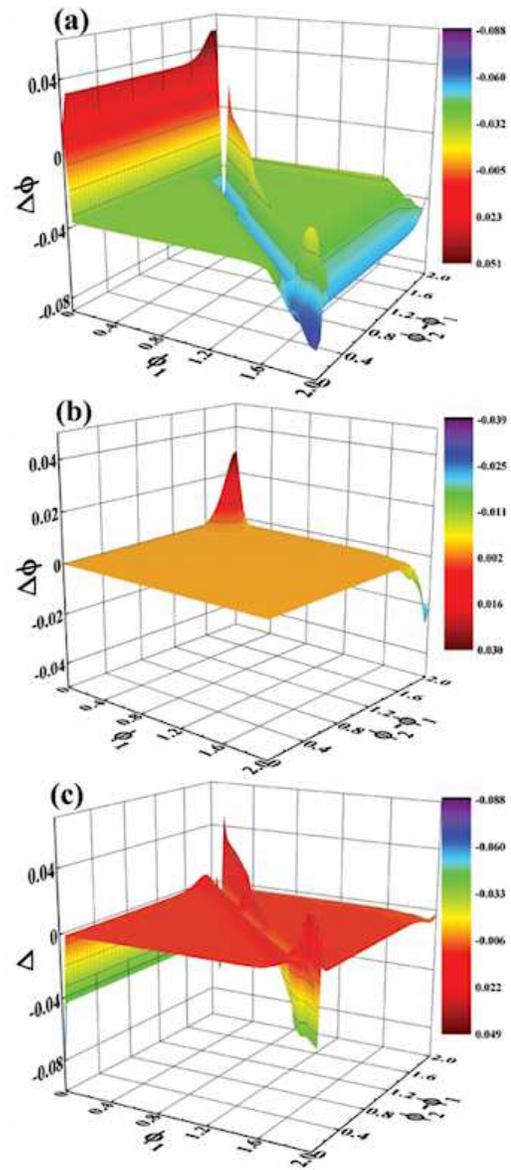}
\caption{(Color online) (a) and (b) respectively show the $PRC$s of first and second order when a mono-block fault is subjected to two successive pulse perturbations. The first perturbation is introduced at the phase $\phi_1$, and the second one is applied with the phase difference $\phi_2-\phi_1$. (c) ilustrates the dependence of the deviation from the superposition principle $\Delta$ on $\phi_1$ and $\phi_2-\phi_1$. The fault parameters are $\epsilon=1.45,\xi=0.5,\gamma=1000$.
\label{Fig6}}
\end{figure}

The first-order $PRC$ is illustrated in Fig. \ref{Fig6}(a). Note that the term $PRC$ is preserved for simplicity, though the plot actually shows the dependence of a phase reset in terms of $\phi_1$ and $\phi_2-\phi_1$. The same terminology is applied when describing the analogous three-dimensional plots in the remaining part of the paper.  Regarding Fig. \ref{Fig6}(a), two points on advancing the phase of the seismic cycle stand out. First, if the initial pulse is applied in a narrow interval sufficiently close to the last seismic event ($\phi=0$), the fault's phase is substantially advanced, irrespective of the precise point when the second perturbation occurs. Also note that the advancing effect of two pulses is significantly stronger than that of a single pulse, cf. Fig. \ref{Fig4}. The second point refers to the domain of $\phi_1$ values away from the characteristic event. There, the earlier arrival of the first perturbation typically requires a late arrival of the second perturbation in order to cause a substantial phase advancement. However, for sufficiently large $\phi_1$, the phase of seismic cycle is advanced only within a narrow interval of preferred $\phi_2$ values, such that the second pulse arrives in a relatively close succession to the first one. Outside of the $(\phi_1,\phi_2)$ domains mentioned above, the impact of two successive pulse perturbations is such that they delay the next characteristic event, viz. the perturbations have a stabilizing effect on the fault.

In case of the second-order $PRC$, see Fig. \ref{Fig6}(b), one notes a sizeable long-term effect if the first pulse arrives early (small $\phi_1$), and the second pulse is introduced sufficiently late within the given cycle (large $\phi_2-\phi_1$). It is interesting that the long-term effect may lead either to fault destabilization (advanced phase of oscillation) or fault stabilization (delayed phase of oscillation), which depends sensitively on the phase of the second pulse. Note that we interpret phase advancement (retardation) of the seismic cycle as destabilization (stabilization) of the fault because its next characteristic event is precipitated (postponed) by the perturbation. The presence of both types of behavior is quite distinct from the case of a single pulse perturbation, cf. Fig. \ref{Fig4}(b), where the only pronounced effect consists in delaying the next oscillation cycle.

Let us now address the deviation from the superposition principle, which is a nonlinear effect that generally occurs for oscillators with more than one degree of freedom. If the superposition principle were to hold, the phase shift caused by two successive pulses would be given by the sum of the two corresponding $PRC$s, as in case of one-dimensional oscillators. Nevertheless, the actual total phase shift caused by the two pulses in systems with dimension larger than $1$ does not coincide with the linear sum of two $PRC$s and may be derived using the formalism from the beginning of this section. In particular, after the first pulse introduced at the moment $t_p$, the system is reset to the state given by \eqref{eq4}. Just before the second pulse, which arrives at the moment $t_p+\Delta t$, the system's state is $\phi(t_p+\Delta t)=\phi_{+}(t_p)+\omega \Delta t, \textbf{a}(t_p+\Delta t)=\Lambda^{\Delta t}(t_p)\textbf{a}_{+}(t_p)=\Lambda^{\Delta t}(t_p))=\kappa A(0,\phi(t_p),\kappa)$, where $\Lambda^{\Delta t}$ is the appropriate evolution operator for the amplitude. Just after the second pulse, the system's phase is reset to $\phi_{+}(t_p+\Delta t)=\phi(t_p+\Delta)+\kappa \Phi(\textbf{a}(t_p+\Delta t),\phi(t_p+\Delta t),\kappa)$, such that the total phase shift induced by the two successive pulses amounts to \cite{KBP13}
\begin{equation}
\Delta \phi=\kappa Z(\phi(t_p),\kappa)+\kappa \Phi(\textbf{a}(t_p+\Delta t),\phi(t_p+\Delta t),\kappa). \label{eq6}
\end{equation}
An important point is that \eqref{eq6} involves the reset function $\Phi$ which depends on the system's amplitude. The presence of such dependence has been demonstrated to be the reason behind the deviation from the superposition principle for oscillators with more than one degree of freedom. Comparing \eqref{eq6} with the two-pulse $PRC$ for a one-dimensional oscillator, one may obtain an explicit expression for the deviation from the superposition principle. In particular, the total phase shift due to two successive pulses for a one-dimensional oscillator is given by $\delta \phi=\kappa Z(\phi(t_p),\kappa)+\kappa Z(\phi(t_p)+\omega \Delta t+\kappa Z(\phi(t_p),\kappa),\kappa)$, such that the correction term $\Delta=\Delta \phi-\delta \phi$ is equal to
\begin{align}
\Delta&=\kappa \Phi(\Lambda^{\Delta t}\kappa A(0,\phi(t_p),\kappa),
\textbf{a}(t_p),\phi(t_p)+\omega \Delta t \nonumber \\
&+\kappa Z(\phi(t_p),\kappa),\kappa)- \kappa Z(\phi(t_p)+\omega \Delta t+\kappa Z(\phi(t_p),\kappa),\kappa).\label{eq7}
\end{align}

We have numerically determined the correction term $\Delta$ for the fault dynamics described by \eqref{eq1}. The plot illustrated in Fig. \ref{Fig6}(c) indicates that the deviation from the superposition principle is most pronounced in the $(\phi_1,\phi_2)$ domains which admit the advance of phase of the oscillation cycle. In other words, these are the parameter domains where the nonlinear character of system \eqref{eq1} is manifested the most.

\section{Phase response of complex faults} \label{sec:compfaults}

This section concerns the behavior of complex faults, which may in general consist of multiple segments with different elastic and frictional properties. The focus lies with the paradigmatic case of a complex fault made up of two blocks. We first analyze the sensitivity to a perturbation for the homogeneous fault comprised of identical blocks, and then consider the heterogeneous fault, where the blocks are characterized by distinct $\epsilon$ values.

\subsection{$PRC$s for the fault comprised of two identical blocks} \label{identical}

For the homogeneous complex fault, we analyze the scenario where the perturbation on block $1$ acts at phase $\phi_1$ of its oscillation cycle, whereas block $2$ receives a kick with the phase difference $\phi_2-\phi_1>0$. The form of perturbation on both blocks is assumed to be identical.

\begin{figure*}[t]
\centering
\includegraphics[scale=0.65]{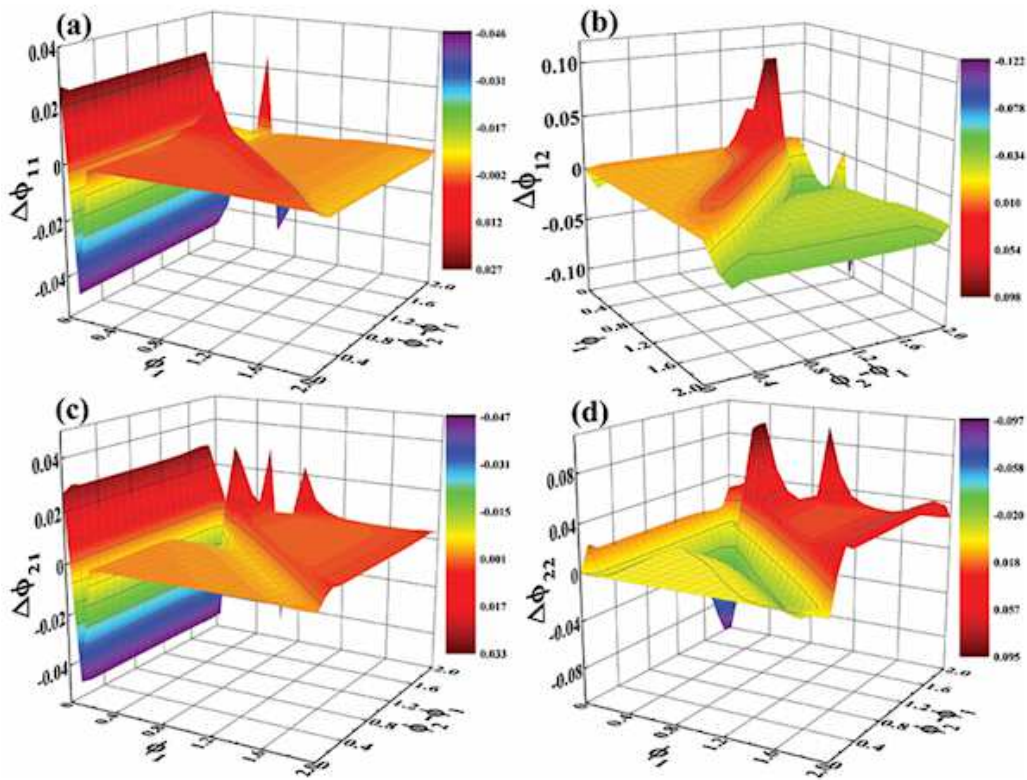}
\caption{(Color online) The top (bottom) row shows the phase responses of the first (left column) and second order (right column) for block $1$ ($2$) in dependence of $\phi_1$ and $\phi_2-\phi_1$. The blocks are assumed to be identical, and are characterized by parameters $\epsilon=1.45,\xi=0.5,\gamma=1000$. The interaction strength $c=0.1$ lies well below the critical bifurcation value and warrants that the periodic oscillations on the coupled blocks are not substantially different from those in the uncoupled case.
\label{Fig7}}
\end{figure*}

The first- and second-order $PRC$s for the appropriate version of system \eqref{eq2} are illustrated in Fig. \ref{Fig7}. The respective phase shifts are denoted by $\Delta\phi_{ij}$, where the first index refers to the particular block, and the second index points to the first/second order of the phase response. It is interesting to compare the first-order $PRC$s in Fig. \ref{Fig7}(a) and Fig. \ref{Fig7}(c), because this indicates how the interaction affects the response of individual blocks. In particular, the phase of both blocks is significantly advanced if the first block is stimulated immediately after the characteristic event. In this case, a perturbation of the first block induces a strong destabilization effect on the dynamics of the second block, irrespective of when the second block is perturbed. Just beyond the described region of $\phi_1$ values, one encounters a narrow domain where the external stimuli delay the cycles of both blocks. Nevertheless, the most interesting point concerns the differences between $\Delta \phi_{11}$ and $\Delta \phi_{21}$ dependences. We find that $\Delta \phi_{21}$ shows a much larger $(\phi_1,\phi_2)$ domain where the phase of the cycle is strongly advanced compared to $\Delta \phi_{11}$. This point corroborates that the dynamics of block $2$ is substantially affected by the perturbation of block $1$ conveyed via the interaction term.  In fact, within the indicated $(\phi_1,\phi_2)$ domain, the destabilization effect on block $2$ is amplified by the coaction of two pulses, reflected in an indirect influence of a perturbation applied to the first block, and a direct impact of the subsequent pulse. Note that the destabilization effect on block $2$ is more pronounced if the perturbation on block $1$ arrives by the end of its oscillation cycle.

As far as the second order $PRC$s are concerned, Fig. \ref{Fig7}(b) and  Fig. \ref{Fig7}(d) both show quite large $(\phi_1,\phi_2)$  domains of substantial phase advancement and phase retardation. These long-term effects are caused by the interaction between the blocks. Note that for the same $(\phi_1,\phi_2)$ values, the long-term effects on two blocks are of different nature. In particular, stabilization (phase delay) of one block is accompanied by a destabilization (phase advancement) of the other block.

\subsection{$PRC$s for the two-block inhomogeneous fault}\label{disparate}

In this subsection, we examine the $PRC$s of an inhomogeneous fault made up of two blocks with disparate $\epsilon$ values. The latter are selected so that the respective oscillation periods of coupled units are quite distinct, $T_1\approx51$ vs $T_2\approx77$. Two different cases are considered: in the first instance, the perturbation is applied only to the block with the shorter oscillation period (here block $1$), whereas in the second instance the block with the longer oscillation period is stimulated (here block $2$). The simulations are carried out in such a way that at the moment when the stimulus arrives to one block, the other block always has the same phase.

\begin{figure*}
\centering
\includegraphics[scale=0.55]{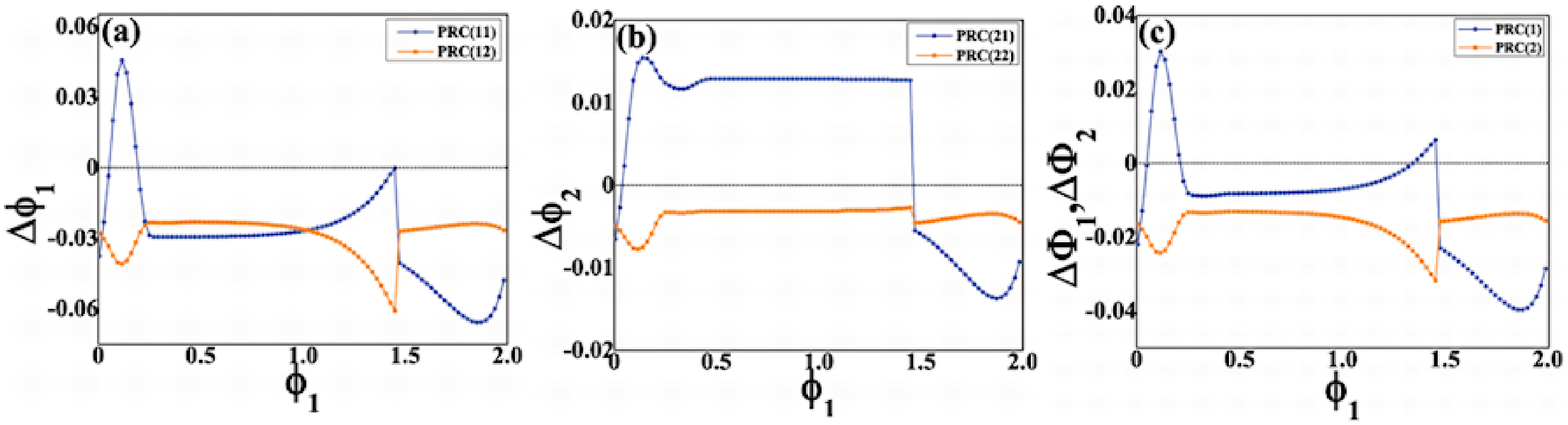}
\caption{(Color online) Scenario where pulse perturbation is introduced to the block with shorter oscillation period.  In (a) are shown the first- (blue circles) and second-order $PRC$s (orange squares) for block $1$, which is subjected to pulse perturbation. (b) illustrates the first- and second-order $PRC$s for block $2$ which is influenced by perturbation only via interaction with block $1$. (c) provides an indication on the average phase response $\Delta \Phi_i, i\in\{1,2\}$ for the total system, viz. the complex fault, whereby index $i$ refers to the first- or second-order dependence.  The block parameters kept fixed are $\xi=0.5,\gamma=1000$, whereas $\epsilon$ values on particular blocks are $\epsilon_1=1.4$ and $\epsilon_2=2$.
\label{Fig8}}
\end{figure*}

The results for the first scenario (perturbation applied to block $1$ at phase $\phi_1$) are illustrated in Fig. \ref{Fig8}, whereby Fig. \ref{Fig8}(a) and Fig. \ref{Fig8}(b) refer to first- and second-order responses of blocks $1$ and $2$, respectively. Note that Fig. \ref{Fig8}(c) shows the average responses
$\Delta \Phi_{i}=(\Delta\phi_{1,i}+\Delta\phi_{2,i})/2$ for the total system (complex fault), where $i\in\{1,2\}$ stands for the first- or second-order response.

As to be expected, for block $1$, the first- and second-order $PRC$s are qualitatively similar to that of an uncoupled block, cf. Fig. \ref{Fig4}(a). As far as block $2$ is concerned, note that Fig. \ref{Fig8}(b) shows the dependence $\Delta \phi_2(\phi_1)$, which is obtained for a fixed value of the phase of the second block. In other words, a perturbation is applied at different phases of the cycle of block $1$, whereas block $2$ at the moment of pulse arrival to block $1$ always lies at a certain fixed phase $\phi_2$. The first-order response of block $2$ implies that the interaction may play an important role in destabilization of the fault. In particular, a perturbation acting on the block with a shorter oscillation period (block $1$) is found to substantially advance the phase of the block with the longer oscillation period (block $2$) for a broad interval of $\phi_1$ values. Note that Fig. \ref{Fig8}(c) implies that the average response of the two-block system is dominated by the behavior of block $1$ where the perturbation is actually applied.

Now let us consider the case of an inhomogeneous two-block fault model where the block characterized by the longer oscillation period (block $2$) is perturbed. In analogy to the case above, a perturbation is applied at different phases of the cycle of block $2$, whereas block $1$ at the moment of pulse arrival to block $2$ always has a fixed phase value $\phi_1$. The first-order responses of the blocks are shown in Fig. \ref{Fig9}(a), whereas Fig. \ref{Fig9}(b) refers to the second-order responses. The average first- and second-order response of the complex two-block fault is provided in Fig. \ref{Fig9}(c).

\begin{figure*}[t]
\centering
\includegraphics[scale=0.55]{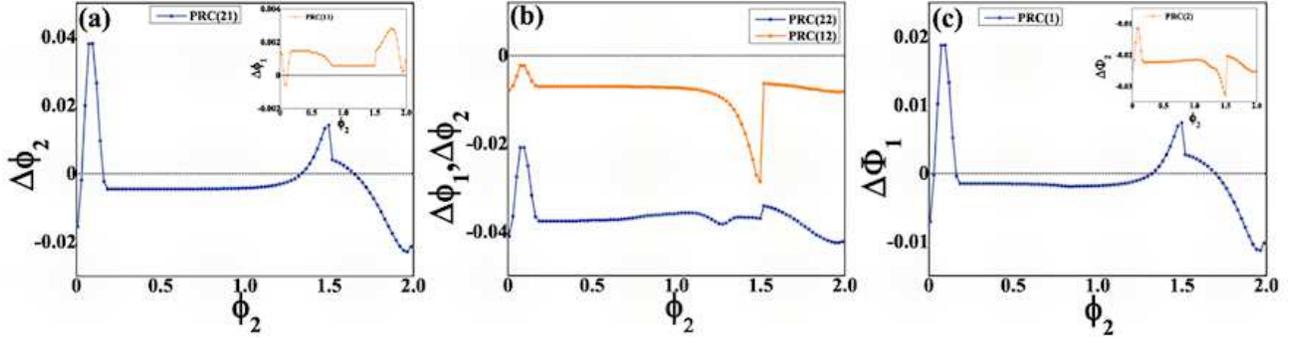}
\caption{(Color online) Scenario where pulse perturbation acts on the block with longer oscillation period.  The main frame and inset in (a) shows the first-order $PRC$ for block $2$ and block $1$, respectively. In (b) are shown the second-order $PRC$s, whereby the blue circles (orange squares) are reserved for block $2$ (block $1$). The main frame and the inset in (c) illustrate the average first- and second-order phase response for the complex fault, respectively. The block parameters are the same as in Fig. \ref{Fig8}.\label{Fig9}}
\end{figure*}

At variance with the scenario considered in Fig. \ref{Fig8}, the first-order $PRC$ for the kicked block now shows two phase intervals which admit an advancement of the seismic cycle, one closely after the characteristic event ($\phi_2\approx 0.1\pi$), and the other located by the end of the seismic cycle ($\phi_2\approx 1.5$). Nevertheless, an important qualitative finding on the first-order response of block $1$ is that for almost all $\phi_2$, the perturbation on block $2$ advances the oscillation cycle on block $1$. The analogous effect of phase advance has already been seen in Fig. \ref{Fig8}(b), but not in such a broad domain of perturbation phases. As far as the total system is concerned, the first-order response is mostly influenced by the behavior of the block explicitly affected by the perturbation, whereas the leading delay effect in the second-order response derives from the other block, cf. Fig. \ref{Fig9}(c).

\section{Summary}\label{sec:summ}

In this paper, we have used the framework of $PRC$s to analyze effects of external perturbations in basic models of earthquake fault dynamics. To our knowledge, such an analysis has not been applied earlier in this field, but has been successfully implemented in the fields of neuroscience and the general theory of systems of coupled phase oscillators. The considered models qualitatively reproduce the stick-slip behavior typical for earthquake motion. Nevertheless, the very notion that the fault dynamics resembles to a relaxation oscillator cannot hold in general, but may be considered as a first approximation to behavior of faults which exhibit characteristic earthquakes with a well-defined recurrence period and low variability (the comparably small coefficient of variation for the timing of the events). Within the proposed concept, external perturbations can influence the duration of the seismic cycle where they have occurred, and may also result in long-term effects, reflected in a change of the subsequent oscillation period. These two points are qualitatively illustrated by the profiles of the obtained first and second-order $PRC$s, respectively. The impact of perturbations can be interpreted as either stabilizing or destabilizing to fault dynamics, in a sense that the external stimuli may either advance the phase of the seismic cycle, thereby precipitating the next characteristic event, or may delay the cycle, thus postponing the next large event.

Our study has been concerned with the models of a simple mono-block fault, as well as paradigmatic examples of complex faults involving two identical or distinct blocks. For a mono-block fault, we have examined how the underlying dynamics is affected by a single or two successive pulse perturbations. In the former case, it is found that external stimuli typically delay the phase of the given oscillation cycle. The exception to this behavior is provided by the stimuli arriving within a narrow interval just after the characteristic event, which result in advancing the phase of the seismic cycle. The second-order $PRC$s indicate an interesting delaying long-term effect for pulses that arrive by the end of the given cycle, which is likely associated with a strong logarithmic nonlinearity of the underlying model. The obtained $PRC$ profiles are shown to be relatively robust to variation of fault parameters. The fault dynamics under the influence of two successive pulses is more complex, and involves two different mechanisms that may give rise to phase advancement. One mechanism is dominated by the first pulse and is completely analogous to what is found in case of a single perturbation, but the other mechanism is qualitatively distinct and requires that the pulses arrive with a specific phase difference.

For a homogeneous two-block fault, we have considered the scenario where each block is affected by a single pulse perturbation. This is realized by selecting a block which is always perturbed before the other block. The first-order $PRC$s indicate that the most likely outcome is fault destabilization, viz. the advance of oscillation phase at both blocks. Such a behavior is contributed by the interaction between the blocks. The second-order $PRC$s reveal highly complex long-term effects, which may be stabilizing or destabilizing to fault dynamics, depending on the times of pulse arrivals. It is interesting that the long-term effects on the blocks can be asymmetric, in a sense that the cycle of one block is advanced, whereas the cycle of the other block is delayed.

For a heterogeneous two-block fault, we have examined scenarios where the block with a shorter or a longer oscillation period receives a single pulse perturbation. In both instances, the simulations are carried out in such a way that at the moment when the stimulus arrives to one block, the other block always has the same phase. An interesting point concerns the advance of phase displayed by the first-order $PRC$s of the respective blocks that are not subjected to pulse perturbation. It turns out that the effect of perturbation conveyed via interaction between the blocks is non-negligible, and its impact on the block that has not received the pulse perturbation is found to be typically destabilizing.

One should caution that the results obtained here cannot be considered within the context of earthquake hazard assessment, nor can immediately be tied to studies of the earthquake triggering effect \cite{KFPVT14,B03,F05}. In reference to the latter point, an interesting issue would be to examine the sensitivity of faults to a stronger perturbation that may affect the amplitude of oscillations. An elaborate investigation of a potential relation between responses of a fault to small perturbation, relevant to $PRC$ theory, and the sensitivity to finite perturbations possibly associated to triggering effect should be an important topic for a future study. Regarding the possible application of the current results, one notes that at variance with the case of a monoblock fault, the $PRC$s for heterogeneous two-block fault exhibit phase advancement for perturbation acting at the later stages of the cycle, cf. Fig. \ref{Fig8}(b) and Fig. \ref{Fig8}(c), as well as Fig. \ref{Fig9}(a) and \ref{Fig9}(c). At both instances, the advance of phase cycle is found for the block with the longer oscillation period and for the total phase of the compound fault, both in cases where the given block is directly perturbed or when the perturbation is transferred via interaction with the other block. It is reasonable to suggest that the perturbation destabilizing the fault in such a fashion gives rise to a clock advance effect which may be seen as a paradigm for studying the appearance of aftershocks \cite{FB06}.

It would be interesting to determine how generic are the results obtained, i. e. whether the $PRC$ profiles found here can be corroborated for other models of earthquake fault dynamics involving different approximations, containing more complex fault structure and featuring distinct friction laws. In a broader perspective, one wonders whether it would be possible to classify different models of fault dynamics in a fashion similar to what has been done in other fields, e. g. for the neuron models where class I excitability is dominated by phase-advance dynamics, whereas class II excitability has both the regimes of phase advance and delay \cite{T07,SPB12}.

\begin{acknowledgments}
This research was supported by the Ministry of Education, Science and Technological Development of the Republic of Serbia (project Nos. ON171017 and OI1611005 to IF, and project No. ON17176016 to SK), the Slovenian Research Agency (Grants P5-0027 and J1- 7009 to MP), the DFG/FAPESP (Grant IRTG 1740/TRP 2011/50151-0 to JK), the Russian Foundation for Basic Research (Grants 15-32-50402 and 15-02-04245 to VN, Grant 14-02-00042 to VK), the Government of the Russian Federation (Agreement No. 14.Z50.31.0033 to VK, VN and JK) and the Ministry of Education and Science of the Russian Federation (Agreement No. MK-8460.2016.2 to VK). The numerical simulations were run on the PARADOX supercomputing facility at the Scientific Computing Laboratory of the Institute of Physics Belgrade.
\end{acknowledgments}

\end{document}